\documentclass[preprint]{aastex}
\usepackage{natbib}

\begin{document}

\title {Millimeter-wave Searches for Cold Dust and Molecular Gas
around T~Tauri Stars in MBM~12}

\author{Michiel R. Hogerheijde\altaffilmark{1,3},
        Ray Jayawardhana\altaffilmark{2,3},
        Doug Johnstone\altaffilmark{4},\\
        Geoffrey A. Blake\altaffilmark{5,6},
    and Jacqueline E. Kessler\altaffilmark{6}}

\altaffiltext{1}{Steward Observatory, The University of Arizona,
933 N. Cherry Ave, Tucson, AZ 85721, U.S.A.}
\altaffiltext{2}{Department of Astronomy, University of
Michigan, 830 Dennison Building, Ann Arbor, MI 48109, U.S.A.}
\altaffiltext{3}{Department of Astronomy, University of California,
Berkeley, 601 Campbell Hall, Berkeley, CA 94720, U.S.A.}
\altaffiltext{4}{Herzberg Institute of Astrophysics, National
Research Council of Canada, 5071 West Saanich Road, Victoria, BC V9E
2E7, Canada}
\altaffiltext{5}{Division of Geological and Planetary Sciences,
California Institute of Technology 150--21, Pasadena, CA 91125, U.S.A.}
\altaffiltext{6}{Division of Chemistry and Chemical Engineering,
California Institute of Technology, Pasadena, CA 91125, U.S.A.}

\begin{abstract}
We report results of a sensitive search for cold dust and molecular
gas in the disks around 8 T~Tauri stars in the high-latitude cloud
MBM~12.  Interferometric observations of 3~mm continuum emission in 5
fields containing 6 of the objects, and literature values for the
remaining two, limit the disk masses to $M_{\rm
disk}<0.04$--0.09~M$_\odot$ (gas+dust), for a gas:dust mass ratio of
100 and a distance of 275~pc. By coadding the 3~mm data of our five
fields, we set an upper limit to the average disk mass of $\bar M_{\rm
disk}(N=5)<0.03$~M$_\odot$.  Simultaneous observation of the CS
$J$=2--1 and the N$_2$H$^+$ 1--0 lines show no emission. Single-dish
observations of the $^{13}$CO 2--1 line limit the disk mass to
(5--10)$\times 10^{-4}$ M$_\odot$ for a standard CO abundance of
$2\times 10^{-4}$. Depletion of CO by up to two orders of magnitude,
through freezing out or photodissociation, can reconcile these
limits. These mass limits lie within the range found in the
Taurus-Auriga and $\rho$~Oph star-forming regions
(0.001--0.3~M$_\odot$), and preclude conclusions about possible
decrease in disk mass over the 1--2 Myr age range spanned by the
latter two regions and MBM~12. Our observations can exclude the
presence in MBM~12 of T~Tauri stars with relatively bright and massive
disks such as T~Tau, DG~Tau, and GG~Tau.
\end{abstract}


\section{Introduction\label{s:intro}}

Characterizing the disks of young stars is an outstanding challenge in
star- and planet-formation research. Millimeter emission from cold
dust and molecular gas has revealed rotating disks a few hundred AU in
radius (\citealt{mannings:haebe,mannings:haebe2};
\citealt{dutrey:tts13co,dutrey:gmaur}; \citealt*{dutrey:dmtau+ggtau};
\citealt*{simon:ttsmass}) around T~Tauri and Herbig Ae/Be stars, while
millimeter and infrared observations have shown the mass in small dust
particles to decrease gradually from the classical-T~Tauri to late
pre--main-sequence stage (\citealt*{robberto:sf99};
\citealt{rayjay:mirtwa, carpenter:ic348}). Evidence suggests that the
age range of 1--3~Myr is pivotal in the disk's evolution, when small
dust in the inner disk regions starts to clear out
\citep{haisch:freq_clusters, luhman:mbm12, hartmann:agespread}. Some
of the dust may have coagulated into larger, centimeter-sized objects,
which are invisible to (sub-) millimeter or infrared observations. The
fate of the gas that makes up 99\% of the disk's mass is only now
being investigated
\citep{thi:h2_ggtau,duvert:exdisk,thi:h2_debris,thi:h2_co_disks,herczeg:h2twa,richter:h2disk}.
This paper investigates the dust- and gas-content of the disks around
several young stars in the MBM~12 cloud.

The cloud MBM~12 (distance 275~pc; but see below) is only one of two
high-latitude clouds known to harbor young stellar objects (YSOs); the
other being MBM~20 \citep*{sandell:l1642}. \citet{hearty:mbm12+20}
identified 8 late-type young stars in MBM~12, and two main-sequence
stars with unclear relation to the cloud. \citet{luhman:mbm12} found
four additional pre--main-sequence members from 2MASS data, and
estimated the cluster's age at
$2^{+3}_{-1}$~Myr. \citet{rayjay:mbm12_mir} detected mid-infrared
excess toward six of the original eight objects. Based on their
H$\alpha$ equivalent line widths, these objects are classified as six
classical T~Tauri stars and two `weak-line' T~Tauri stars (Table
\ref{t:coords} and Fig. \ref{f:map}).  Their frequency of K- (20\%)
and L-band infrared excesses (70\%) suggests clearing of
their disks \citep{luhman:mbm12}. A crucial but unresolved question is
how much material, if any, still resides in more extended, outer
disks. Thus far, upper limits to the millimeter continuum of only two
of the T~Tauri stars (E02553+2018 and LkH$\alpha$~264) have been
reported \citep{pound:mbm12}.

The group of young stars in MBM~12 is among several nearby `isolated'
groups that have recently been identified and investigated for the
nature of their disks \citep{rayjay:newblock,rayjay:youngstars}. Some
of these groups (the TW~Hydrae Association (TWA) at $\sim 55$~pc,
\citealt{kastner:twa}; and $\eta$ Chamaeleontis cluster at $\sim 97$~pc,
\citealt*{mamajek:etacha}) are far from obvious parent molecular
clouds. This suggests ages for these stars of 5--10~Myr, and little
circumstellar disk material is expected. The association of the MBM~12
stars with cloud material supports their younger derived age, and they
may be represent an earlier epoch of groups like TWA and $\eta$~Cha.

Until recently, MBM~12 was thought to be the nearest star-forming
cloud to the Sun, at $\sim 65$~pc \citep{hearty:mbm12_rosat}. New
evidence suggests it may be significantly further
\citep{luhman:mbm12}, at $\sim 275$~pc. \citet*{idzi:mbm12_aas} found
extinction at 65, 140 and 275~pc along the MBM~12 line of sight, so
the distance of the stars is unclear. We follow \citet{luhman:mbm12}
and adopt 275~pc, which gives the most plausible location of the stars
on the the Herzsprung-Russel Diagram. In any case, the young age
indicated by MBM~12's association with cloud material and its relative
proximity make it an excellent target to study circumstellar disks in
detail.

This paper presents the results of sensitive millimeter-wave searches
for cold dust and molecular gas associated with the six classical
T~Tauri stars in MBM~12 known to harbor disks for which no millimeter
data exist, and the two weak-line T~Tauri stars (Table
\ref{t:coords}). We did not include the newly identified members of
MBM~12 \citep{luhman:mbm12}, but fortuitously include the edge-on disk
source MBM~12~3C \citep{rayjay:mbm12_ao} which happened to fall in our
field containing LkH$\alpha$~262 and LkH$\alpha$~263. Since the
completion of these observations, \citet{luhman:mbm12} has concluded
that one object, RXJ0306.5+1921, is likely an older interloper. For
completeness, we still report the data on this object.  We used a
millimeter interferometer to measure the continuum emission, because
the high spatial resolution avoids confusion with the surrounding
cloud. We use single-dish submillimeter observations of $^{13}$CO
$J$=2--1 to trace any cold ($\sim 16$ K) and relatively low density
($n_{\rm H_2}\approx 10^4$ cm$^{-3}$) gas that may reside around the
objects. Section \ref{s:obs} describes the observations and section
\ref{s:results} the resulting mass limits. Section \ref{s:discussion}
discusses these limits in the context of other nearby star-forming
regions and young associations. Section \ref{s:summary} concludes the
paper with a short summary.

\section{Observations\label{s:obs}}

\subsection{Owens Valley Radio Observatory\label{ss:ovro}}

The 3~mm continuum observations were carried out with the Millimeter
Array at the Owens Valley Radio Observatory (OVRO)\footnote{The Owens
Valley Millimeter Array is operated by the California Institute of
Technology under funding from the U.S.\ National Science Foundation
(\#AST96--13717).} on 2001 March 19, 20, and 24. The receivers were
tuned to the frequencies of C$^{34}$S $J$=2--1 at 96.41294~GHz and the
N$_2$H$^+$ 1--0 triplet at 93.17370~GHz ($\lambda=3$~mm), where their
performance is optimum. The lines were recorded in two 7.75~MHz-wide
bands of 62 channels, while the continuum signal in the upper and
lower sidebands was recorded over a total bandwidth of 2~GHz. The
array was in the `E' configuration, with baselines between 10 and
40~k$\lambda$ resulting in a synthesized beam of $5''$. In each
eight-hour track two sources were observed by alternating between them
every half hour. Source coordinates are listed in Table
\ref{t:coords}. LkH$\alpha$~262 and 263 are located $17''$ from each
other, well within the $64''$ primary beam of the 10-m antennas; they
were observed in a single pointing for an entire eight-hour track. The
edge-on disk source MBM~12A~3C \citep{rayjay:mbm12_ao} is located
$\sim 4''$ from LkH$\alpha$~263 and also falls within the primary
beam. The complex gain was calibrated by observations every 30 minutes
of the nearby ($9^\circ$) quasar 0235+164 (flux 1.0~Jy). The phase
stability was less than optimal, and only data with coherences above
$\sim 60$\% were used.

Flux scaling was obtained from observations of Uranus and the quasar
3C84 (flux 6.2~Jy, bootstrapped from the Uranus data). The data were
calibrated in the MMA software package \citep{scoville:database}, and
further processed with the MIRIAD package \citep*{sault:miriad}. The
resulting 1$\sigma$ rms noise level of the side-band averaged
continuum was 1.3--2.9~mJy in the $\sim 5''$ beam. We also coadded the
observed five fields, aligning the source positions
(Fig. \ref{f:coadd}), to derive constraints on the `average' disk mass
of the six objects (see \citealt{carpenter:ic348}).

\subsection{James Clerk Maxwell Telescope\label{ss:jcmt}}

The $^{13}$CO $J$=2--1 observations were carried out at the James
Clerk Maxwell Telescope (JCMT)\footnote{ The JCMT is operated by the
Joint Astronomy Centre in Hilo, Hawaii on behalf of the parent
organizations Particle Physics and Astronomy Research Council in the
United Kingdom, the National Research Council of Canada and The
Netherlands Organization for Scientific Research.} in Canadian service
mode on 2001 March 23--25, April 9, May 24, and May 25. The facility
receiver `RxA3' was tuned to the line frequency at 220.3986765~GHz and
the spectra were recorded in Digital Autocorrelation Spectrometer
backend in a bandwidth of 125~MHz and a resolution of 78~kHz
(0.11~km~s$^{-1}$). Typical system temperatures of 400--600~K were
found, giving a noise rms of 67--100~mK per channel after on-source
integration of 20 minutes; a main beam efficiency of 0.6 has been
applied. The observations were carried out in a frequency switched
mode with a throw of 8.2~MHz to compensate for atmospheric
contributions. The data were analysed with the SPECX software package.

The observations were pointed at the source positions of
Table~\ref{t:coords}; S18 was not observed due to time constraints. If
emission was detected at that position, an integration at an offset
position of $(+30'',0'')$ was taken. If no emission at $(0'',0'')$ was
detected, or if the emission at both positions was equal within the
error, we concluded that there was no molecular gas associated with
the source; in the latter case we attributed all emission to the
surrounding cloud.

\section{Results\label{s:results}}

\subsection{Continuum emission from OVRO\label{ss:continuum}}

The OVRO observations did not detect 3~mm continuum
emission, to 3$\sigma$ upper limits of 4--9~mJy; for LkH$\alpha$~264
and E02533+2018 \citet{pound:mbm12} reported upper limits at 2.7~mm
with a 3$\sigma$ level of 15 and 11~mJy, respectively (Table
\ref{t:cont}). No C$^{34}$S or N$_2$H$^+$ emission was detected in the
OVRO data, at a 3$\sigma$ noise level of $\sim 0.5$ Jy~beam$^{-1}$
(3~K) in 0.40 km~s$^{-1}$ channels. These upper limits on the line
emission do not provide useful constraints, and we will not analyse
them further.

We derive upper limits on the mass traced by the continuum emission
using the simplifying assumptions that the emission is optically thin
and characterized by a single dust temperature and emission
coefficient (see, e.g., \citealt{beckwith:mmdisks,duvert:exdisk}).
In the Rayleigh-Jeans limit valid for the observing wavelength and
typical dust temperatures, the disk mass is related to the continuum
flux as%
\begin{equation}
M_{\rm disk} =
   0.01 M_\odot 
   \Bigl({{F_\nu(3\,{\rm mm})}\over{1\ {\rm mJy}}}\Bigr)
   \Bigl({{D}\over{275\ {\rm pc}}}\Bigr)^2
   \Bigl({{\kappa}\over{0.6\ {\rm cm^2\ g^{-1}_{dust}}}}\Bigr)^{-1}
   \Bigl({{T_{\rm dust}}\over{25\ {\rm K}}}\Bigr)^{-1}.
   \label{eq:dust2}
\end{equation}
Here, $M_{\rm disk}$ is the disk mass, $F_\nu(3\,{\rm mm})$ is the
flux, $D$ is the distance, $\kappa$ the emissivity per unit dust mass,
and $T_{\rm dust}$ the dust temperature. $M_{\rm disk}$ refers to the
mass of the disk in gas and dust. We have adopted a gas:dust mass
ratio of 100:1 as is found in the translucent interstellar
medium. Since we measure the dust content only, our actual limits are
$1\times 10^{-4}$~M$_\odot$ of dust per 1~mJy of 3~mm flux. The
adopted emissivity is a value `typical' for evolved dark clouds or
protostellar disks \citep{beckwith:beta,pollack:kappa,
ossenkopf:kappa} and is uncertain by a factor of a few. The dust
temperature is uncertain by less than a factor of 2. We have used the
distance of 275~pc; at 65 or 140~pc the derived mass would be smaller
by factors of 18 and 4, respectively. Table \ref{t:cont} lists the
resulting mass limits of 0.04--0.09 M$_\odot$ (gas+dust). This
includes the values of \citet{pound:mbm12} for E02553+2018 (0.07) and
LkH$\alpha$~264 (0.09 M$_\odot$), recalculated for our adopted dust
properties and temperatures and the larger distance of 275~pc.

We can place a generous upper limit to the disk's diameter by assuming
it is fully optically thick at 3~mm and geometrically thin.  At 3~mm
we find an upper limit of 26~AU for a flux limit of 1~mJy, and at
2.7~mm of 20~AU. The resulting limits for our sample of 50--80~AU are
significantly larger than the areas of circumstellar disks that are
likely to be optically thick; with $\kappa = 0.6$ cm$^{2}$~g$^{-1}$
(dust), a 50~AU diameter fully optically thick ($\tau>3$) disk has a
mass $>0.1$~M$_\odot$ (gas+dust).

We also calculated the continuum flux of a disk using more realistic
descriptions by \citet{chiang:disksed} and
\citet{dalessio:disk1}. Using the former model, scaled directly to
0.01~M$_\odot$ without taking into account any changes to the disk's
internal structure, we find a flux close to that of equation
(\ref{eq:dust2}).

The coadded five fields yield a limit on the `average' disk mass for
the six objects. We find a 1$\sigma$ rms noise level of 1~mJy, and an
average disk mass below $\bar M_{\rm disk}(N=5)<0.03$~M$_\odot$. This
average includes the data on the object RXJ0306.5+1921 that
\citet{luhman:mbm12} conclude may not be a member of MBM~12. If we
exclude this object, the resulting noise rms (1.1 mJy) and $\bar
M_{\rm disk}$ limits (0.033 M$_\odot$) do not change significantly.

\subsection{Line emission from JCMT\label{ss:lines}}

Four of our objects do not show $^{13}$CO 2--1 emission with a
3$\sigma$ value of 0.25~K; the remaining have detected emission in the
range of 0.7--1.3~K, but show similar values at a position $30''$ away
from the star (Table~\ref{t:lines} and Fig. \ref{f:spectra}). Only for
E02553+2018 does an additional component at $V_{\rm LSR}=-1.5$~km~s$^{-1}$
shows up at $(+30'',0'')$, likely associated with the MBM~12 cloud. We
conclude that for the latter three objects all emission comes from the
surrounding cloud, and that the 3$\sigma$ noise level of 0.25~K can be
adopted as upper limit to any gas mass associated with these objects.

Following \citet{scoville:n2071+w49+n7538}, who relate the integrated
intensity to the beam-averaged column density of a linear rotor such
as $^{13}$CO, and using a dipole moment of 0.112~Debye for $^{13}$CO
and a $^{13}$CO/H$_2$ abundance of $3\times 10^{-6}$
\citep{lacy:co_h2, wilson:abundances}, we find a disk (gas) mass of%
\begin{equation}
M_{\rm disk} = (5-8) \times 10^{-4}\ M_\odot 
\Bigl({\rm {[^{12}CO]:[H_2]}\over{2\times 10^{-4}:1}}\Bigr)
\Bigl({\rm {[^{12}CO]:[^{13}CO]}\over{65:1}}\Bigr)
\Bigl({D\over{\rm 275\,pc}}\Bigr)^2
\Bigl({{\int T_{mb} dV}\over{0.27\, {\rm K~km~s^{-1}}}}\Bigr).
\label{eq:gasmass}
\end{equation}
Equation (\ref{eq:gasmass}) assumes that the emission is optically thin
and uses the observed upper limit of 0.3~K per 0.11 km~s$^{-1}$
channel or 0.27~K~km~s$^{-1}$ for an adopted line width of
2~km~s$^{-1}$. This line width is appropriate for disks in Keplerian
rotation around a 0.5~M$_\odot$ star under most inclinations. The
range in masses in equation (\ref{eq:gasmass}) reflects the considered
range excitation temperatures, 6~K~$<T_{\rm ex}<$~70~K. The relation
reaches a minimum at $T_{\rm ex}=16$~K, which corresponds to the
energy of the $J$=2 level. Table \ref{t:lines} lists the resulting
mass limits for our sample of (5--10)$\times 10^{-4}$~M$_\odot$.

Using the \citet{chiang:disksed} model, chemical calculations from
\citet{aikawa:chem2d}, and adopting varying disk inclinations, we
confirm again that a more realistic treatment of the density and
temperature structure gives results consistent with equation
(\ref{eq:gasmass}).

\section{Discussion\label{s:discussion}}

Section \ref{s:results} places upper limits on the mass of cold dust
and gas associated with eight T~Tauri stars in MBM~12. The 3~mm
continuum limits the mass to $<0.04$--0.09~M$_\odot$. The $^{13}$CO
lines provide limits that are much lower, (5--10)$\times
10^{-4}$~M$_\odot$. However, the abundance of molecules such as CO may
be decreased as commonly observed in T~Tauri disks (e.g.,
\citealt*{dutrey:ggtau}; \citealt{dutrey:tts13co}), possibly by
freezing out onto dust grains and by photodissociation by (inter-)
stellar ultraviolet photons as suggested by chemical models (e.g.,
\citealt{aikawa:chem2d}; \citealt{willacy:photodisk}). These processes
can easily reduce the CO abundance by factors of tens or a hundred, in
which case the limits of Table \ref{t:lines} are no longer lower than
those of Table \ref{t:cont}. We stress that if CO is significantly
frozen out, the gas mass of the disk does not change appreciably,
since it is dominated by undepleted H$_2$. If CO is significantly
photodissociated, the total molecular gas mass may be reduced if a
sizable fraction of H$_2$ is also photodissociated.

\citet{rayjay:mbm12_mir} report N-band excess in the six classical
T~Tauri stars, indicating the presence of material close to the
stars. This places a lower limit to the disk mass of $\sim
10^{-5}$~M$_\odot$, well below our upper limits.  The edge-on disk
source MBM~12A~3C near LkH$\alpha$~263 \citep{rayjay:mbm12_ao}
requires a mass of $\sim 2\times 10^{-3}$~M$_\odot$ to explain its
scattered light image. This mass is comfortably bracketed by our lower
and upper limits. \citeauthor{rayjay:mbm12_ao} model the scattered
light with a distribution of dust sizes. At the observing wavelength
of 3~mm, our observations are sensitive to particles much larger than
those doing the infrared scattering, but both populations are
connected through the adopted dust-size distribution and mass
emissivity coefficients, and therefore refer to a similar mass
reservoir.

With the obtained limits, the disk masses in MBM~12 are
indistinguishable from those found in Taurus-Auriga and $\rho$~Oph of
0.001--0.3~M$_\odot$ \citep{beckwith:mmdisks,
osterloh:mm,andre:rhoophmm}. We therefore cannot draw any conclusions
about the evolution of disks in the 1--2 Myr age range spanned by
these regions and MBM~12. While the work by \citet{luhman:mbm12}
suggests that the dust in the disks in MBM~12 is starting to clear
out, more sensitive measurements of the dust continuum with, e.g.,
SCUBA or SIRTF, are needed to investigate the fate of the colder dust
at larger radii. \citet{carpenter:ic348} limits the disk masses of the
members of the 2~Myr cluster IC~348 to $<0.025$~M$_\odot$ (or
0.002~M$_\odot$ averaged over his 95 sources), somewhat below our
limits on MBM~12.

Our mass limits do exclude the presence of objects such as T~Tau,
GG~Tau, and DG~Tau with large millimeter fluxes
\citep{beckwith:mmdisks}. We would easily have detected these fluxes, at
20--40~mJy when scaled to 275~pc and extrapolated to 3~mm (adopting
spectral indices between 2.5 and 4). But even in Taurus-Auriga these
objects are rare, and given the small number of T~Tauri stars in
MBM~12 ($\sim 10$; \citealt{luhman:mbm12}) no such bright objects
would necessarily be expected.

\section{Summary\label{s:summary}}

We obtained upper limits on the 3~mm continuum flux and $^{13}$CO 2--1
line intensity of eight T~Tauri stars in the MBM~12 region. These
limits constrain the disk masses to $<0.04$--0.09 M$_\odot$
(gas+dust), not inconsistent with the distribution of masses in
slightly younger regions like Taurus-Auriga and $\rho$~Oph ($\sim
1$~Myr vs.\ $\sim 2$~Myr), and consistent with the mass of $2\times
10^{-3}$ M$_\odot$ derived for the edge-on disk of MBM~12A~3C.  We
exclude the presence of objects such as T~Tau, GG~Tau, and DG~Tau with
bright millimeter emission. More sensitive searches with, e.g., SCUBA
and SIRTF, will probe the evolution of the cold disk material at
larger radii at the moment when the inner disks start to clear.

\acknowledgments 

We wish to thank the staff of OVRO and JCMT, and Henry Matthews in
particular, for their outstanding support. Paola D'Alessio, Erik
Mamajek, Michael Meyer, and James Muzerolle are thanked for useful
discussions. The referee provided many useful comments that
significantly improved the paper.  At the University of California,
Berkeley, the research of MRH and RJ was supported by the Miller
Institute for Basic Research in Science. This work was supported in
part by NASA Origins grant NAG5-11905 to RJ.

\newpage




\newpage

\figcaption[]{Positions of the eight MBM~12 T~Tauri stars studied in
this paper (star symbols) superposed on the IRAS 100~$\mu$m emission
(greyscale) and CO $J$=2--1 emission (contours, at 2, 4, 6,
... K~km~s$^{-1}$; \citealt*{dame:co21}). The additional four members
of MBM~12 identified by \citealt{luhman:mbm12} are indicated by open
squares. RXJ0306.5+1921 and S18 both are located $2^\circ$--$3^\circ$
from the main core of MBM~12; \citet{luhman:mbm12} concludes that S18
is a true member of the association while RXJ0306.5+1921 is likely an
older interloper.\label{f:map}}

\figcaption[]{Reconstructed image of the coadded 3~mm continuum data
from OVRO. The five observed fields have been shifted to a common
origin (indicated by the cross), taking into account the source
offsets for the field that contains both LkH$\alpha$ 262 and 263. The
coadded data place an upper limit on the `average' disk mass of
0.03~M$_\odot$ (gas+dust). Contours are drawn at 1$\sigma=1$~mJy. The
synthesized beam size is $4.6''\times 3.7''$ and is shown in the lower
left corner. This image includes the data on the likely interloper
RXJ0306.5+1921; without these data the resulting image is not
significantly different.\label{f:coadd}}

\figcaption[]{Spectra of $^{13}$CO $J$=2--1 obtained with the JCMT
toward seven of our eight objects. Solid lines show the spectra to the
source position. In three cases emission was detected, and spectra at
a nearby ($30''$) cloud position were obtained (dotted lines). In all
three cases the emission was found to be equal, within the noise
level, at both positions.\label{f:spectra}}


\newpage

\begin{deluxetable}{lrrcc}
\tablecaption{Observed Sources\label{t:coords}}
\tablecolumns{5}
\tablehead{
& \colhead{$\alpha$(2000)} & \colhead{$\delta$(2000)} & 
  \colhead{K-L, K-N\tablenotemark{a}}\\
\colhead{Source} & \colhead{(h m s)} & \colhead{($^\circ$ $'$ $''$)} &
  \colhead{Excess?} & 
  \colhead{Classification\tablenotemark{b}}
}
\startdata
RUG0255.4+2005 & 02 55 25.7 & 20 04 53 & no  & wTTs \\
LkH$\alpha$262 & 02 56 07.9 & 20 03 25 & yes & cTTs \\
LkH$\alpha$263 & 02 56 08.4 & 20 03 39 & yes & cTTs \\
LkH$\alpha$264 & 02 56 37.5 & 20 05 38 & yes & cTTs \\
E02553+2018    & 02 58 11.2 & 20 30 04 & yes & c/wTTs\tablenotemark{c} \\
RXJ0258.3+1947 & 02 58 15.9 & 19 47 17 & yes & cTTs \\
RXJ0306.5+1921 & 03 06 33.1 & 19 21 52 & no  & wTTs\tablenotemark{d} \\
S18            & 03 02 21.1 & 17 10 35 & yes & cTTs\tablenotemark{e} \\
\enddata
\tablenotetext{a}{From \citet{rayjay:mbm12_mir}.}
\tablenotetext{b}{Defining classical T~Tauri stars (cTTs) as objects
having H$\alpha$ equivalent linewidths $>5$--10~\AA, and weak-line
T~Tauri stars (wTTs) as EW(H$\alpha$)$<5$--10~\AA.}
\tablenotetext{c}{E02553+2018 has an uncertain classification as
classical or weak-line T~Tauri star: it has detected infrared excess
and EW(H$\alpha$)$\approx 4$~\AA.}
\tablenotetext{d}{\citet{luhman:mbm12} conclude that RXJ0306.5+1921 is
likely not a member of MBM~12, but an older interloper.}
\tablenotetext{e}{In spite of its distance from the core of the MBM~12
cloud, \citet{luhman:mbm12} conclude that S18 is a bona-fide member.}
\end{deluxetable}

\begin{deluxetable}{lcccc}
\tablecaption{Upper Limits on $\lambda$=3~mm Continuum Flux\label{t:cont}}
\tablecolumns{5}
\tablehead{
 & \colhead{3$\sigma$} & \colhead{Beam} & \colhead{Upper Limit} & \colhead{Upper Limit} \\
 & \colhead{Upper Limit} & \colhead{Size} 
 & \colhead{Disk Mass\tablenotemark{a}} 
 & \colhead{Diameter\tablenotemark{b}} \\
\colhead{Source} & \colhead{(mJy)} & \colhead{(arcsec)} & 
  \colhead{(M$_\odot$)} & \colhead{(AU)} }
\startdata
RXJ0255.4+2005           &  4 & 5.2 $\times$ 4.0 & 0.04 & 52 \\
LkH$\alpha$262           &  7 & 5.3 $\times$ 3.5 & 0.07 & 69 \\
LkH$\alpha$263           &  7 & 5.3 $\times$ 3.5 & 0.07 & 69 \\
LkH$\alpha$264 & 15\tablenotemark{c} & 3.5 $\times$ 3.1 & 0.09 & 77 \\
E02553+2018    & 11\tablenotemark{c} & 2.7 $\times$ 1.8 & 0.07 & 66 \\
RXJ0258.3+1947           &  5 & 5.2 $\times$ 4.0 & 0.05 & 58 \\
RXJ0306.5+1921           &  7 & 4.9 $\times$ 3.9 & 0.07 & 69 \\
S18                      &  9 & 5.3 $\times$ 3.9 & 0.09 & 78 \\
Average\tablenotemark{d} &  3 & 4.6 $\times$ 3.7 & 0.03 & 45 \\
\enddata
\tablenotetext{a}{Assuming a distance of $d$=275 pc and a gas-to-dust ratio
  of 100. Disk mass scales as 
  $M_d\propto d^{-2}$ and is independent of inclination in these optically thin
  cases}
\tablenotetext{b}{Minimum diameter of a fully optically thick disk
  ($\tau>3$) with a dust temperature of 25~K that would produce a flux
  comparable to the obtained upper limit.}
\tablenotetext{c}{3$\sigma$ Flux limits at 2.7~mm from
  \citet{pound:mbm12}.}
\tablenotetext{d}{Noise level and derived upper limit on the `average' disk
mass obtained by coadding the 5 fields observed with OVRO, i.e., excluding
LkH$\alpha$~264 and E02553+2018.}
\end{deluxetable}

\begin{deluxetable}{lrrr}
\tablecaption{Observed $^{13}$CO $J$=2--1 Line Intensities\label{t:lines}}
\tablecolumns{4}
\tablehead{
 & \colhead{$T_{mb}$} & \colhead{$T_{mb}$} & \colhead{Upper Limit\tablenotemark{a,b}} \\
 & \colhead{$(0'',0'')$} & \colhead{$(30'',0'')$} & \colhead{Mass} \\
\colhead{Source} & \colhead{(K)} & \colhead{(K)} & \colhead{($10^{-4}$
M$_\odot$)} }
\startdata
RXJ0255.4+2005 & $<0.23$         & \nodata         &  6 \\
LkH$\alpha$262 & $<0.28$         & \nodata         &  7 \\
LkH$\alpha$263 & $<0.40$         & \nodata         &  10 \\
LkH$\alpha$264 & $1.27 \pm 0.13$ & $1.03 \pm 0.1$ &  10 \\
E02553+2018    & $0.92 \pm 0.10$ & $1.08 \pm 0.12$ &  9 \\
RXJ0258.3+1947 & $0.72 \pm 0.10$ & $0.62 \pm 0.08$ &  8 \\
RXJ0306.5+1921 & $<0.22$         & \nodata         &  5 \\
S18            & \nodata         & \nodata         &  \nodata \\
\enddata
\tablenotetext{a}{Assuming a distance of $d$=275 pc. Disk mass scales as
   $M_d\propto d^{-2}$ and is independent of inclination.}
\tablenotetext{b}{Depletion of CO by freezing out onto dust grains and/or
photodissociation by (inter-) stellar ultraviolet photons can easily raise 
these limits by two orders of magnitude.}
\end{deluxetable}


\newpage

\begin{figure}
\figurenum{\ref{f:map}}
\epsscale{1.0}
\plotone{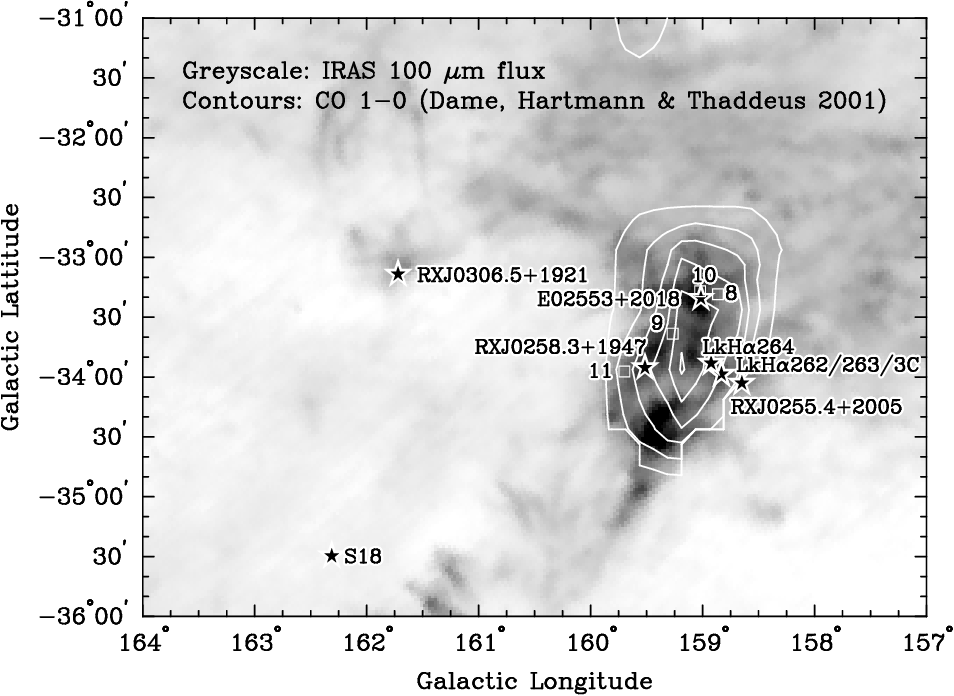}
\caption{}
\end{figure}

\begin{figure}
\figurenum{\ref{f:coadd}}
\epsscale{0.8}
\plotone{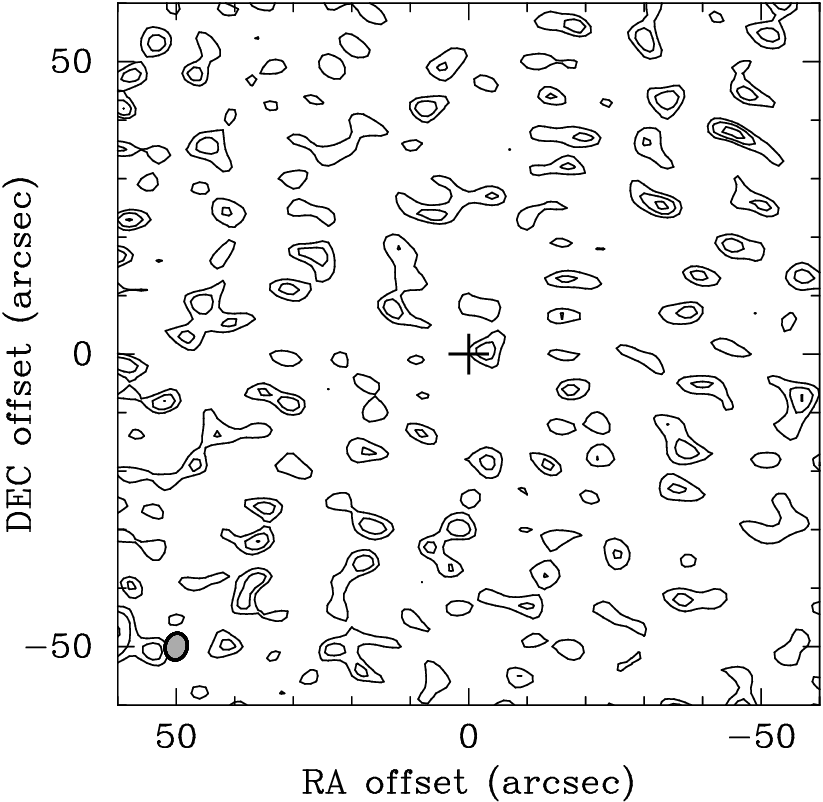}
\caption{}
\end{figure}

\begin{figure}
\figurenum{\ref{f:spectra}}
\epsscale{0.7}
\plotone{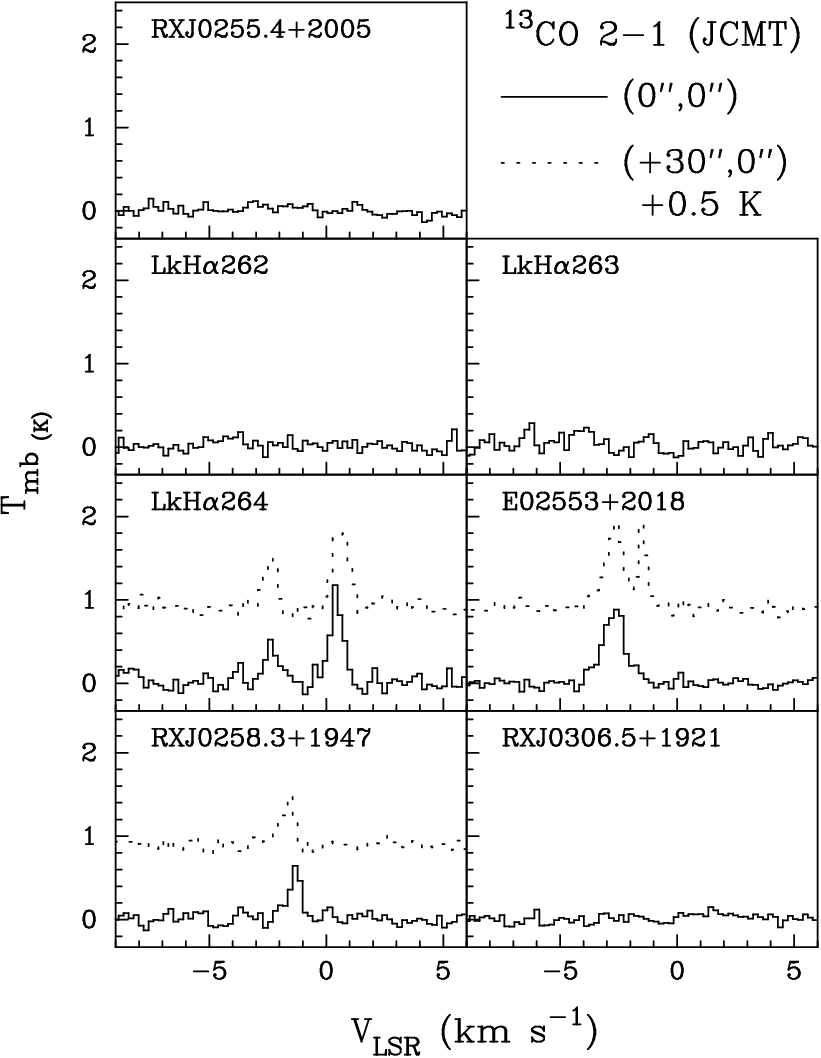}
\caption{}
\end{figure}

\end{document}